\documentstyle[twocolumn,prl,aps,psfig]{revtex}
\begin{document} 

\draft

\wideabs{

\title{
%\Large\bf 
Adatom Diffusion at GaN ($0001$) and ($000\bar1$) Surfaces}
\author{Tosja Zywietz, J{\"o}rg Neugebauer, Matthias Scheffler}
\address{Fritz-Haber-Institut der Max-Planck-Gesellschaft, Faradayweg
4--6, D-14195 Berlin, Germany}
\date{\today}

\maketitle

\begin{abstract}
The diffusion of Ga and N adatoms has been studied for the
technologically relevant wurtzite ($000\bar1$) and ($0001$) surfaces
employing density-functional theory. Our calculations reveal a very different
diffusivity for Ga and N adatoms on the equilibrium surfaces: While Ga
is very mobile at typical growth temperatures, the diffusion of N is
by orders of magnitudes slower. These results give a very detailed insight of how and under which growth
conditions N adatoms can be stabilized and efficiently incorporated at the surface. We further find
that the presence of excess N strongly increases the Ga diffusion barrier and discuss the consequences
for the growth of GaN.
\end{abstract}
}

Recently, great progress in fabricating
highly efficient GaN-based devices has been achieved 
\cite{nakamura94,nakamura97}. Nevertheless, there are still
substantial problems concerning growth optimization and insight
into the fundamental mechanisms is rather shallow. Investigations of molecular beam epitaxy (MBE) growth have shown that the film
structure and morphology are very sensitive to the III/V
ratio\cite{Speck97,brandt96}. GaN films grown under more N-rich
conditions are rough and faceted while going towards more Ga-rich conditions a smoother surface
morphology and better film quality are obtained\cite{Speck97}.

In order to improve growth in a systematic way it is essential to
understand the underlying kinetic processes such as adsorption,
desorption, and surface diffusion. In particular, adatom diffusion on
surfaces is considered to be a key parameter controlling the growth
rate, the material quality, and the surface
morphology. Experimentally, an analysis of surface diffusion is difficult:
So far only {\em effective} diffusion barriers have been obtained for
GaN\cite{brandt96,Weber98}. It is also not clear whether the cation or
anion surface diffusion is the rate limiting process\cite{brandt96}.
A problem in growing GaN is also the high N vapor pressure and considerable 
efforts have been made to enhance the N incorporation. However, the mechanism by which
N is incorporated at the surface could not be identified.
Theoretically, only few studies to describe GaN growth have been
reported, which were either based on thermodynamic
models\cite{Karpov97} or Monte Carlo simulations\cite{Wang94}. These approaches are useful to model growth
on a {\em mesoscopic} scale. However, the microscopic meaning of the effective parameters 
and the underlying {\em microscopic} processes remain unclear. In order to identify the fundamental growth mechanisms,
but also to improve the above mentioned methods, the
correct {\em microscopic} parameters are needed \cite{Ruggerone97}.

We have therefore performed a comprehensive and detailed study of the
migration and energetics of Ga and N adatoms on GaN surfaces,
employing total-energy density-functional theory calculations with {\it ab initio} pseudopotentials. We will focus
on the two technologically relevant orientations for wurtzite GaN, the
(0001) and the (000${\bar 1}$) surfaces. The adatom substrate system
is modeled by supercell geometries with at least 9 layers of
GaN, a 14 bohr vacuum region and ($2\times2$) periodicity. One side of the slab is passivated by fractional
pseudo-hydrogen. Tests revealed that migration paths and
diffusion barriers are not significantly affected by the Ga $3d$ electrons and that
it is sufficient to treat them within the non-linear
core correction (NLCC) \cite{tbpublished}. Details of the calculations are discussed elsewhere
\cite{stumpf94,mrs95_tb_s}.

The energetically stable reconstructions for polar GaN surfaces have
been recently identified for the cubic (001) \cite{neugebauer98} and
the wurtzite ($0001$) and ($000\bar1$) orientations
\cite{smith97}. These studies revealed a surprising feature not
observed for ``traditional'' III-V semiconductor surfaces: Independent
of the chemical environment (Ga- or N-rich) polar GaN surfaces turn
out to be always cation stabilized with no N atoms in or on-top of the
surface layer. The only known exception so far is a N $(2\times2)$
adatom structure on the (0001) surface found to be stable under very N-rich
conditions \cite{smith97}. In order to study adatom diffusion on the {\em
thermodynamical stable surfaces} we will therefore focus on the two structures
shown in Fig. 1 which both are characterized by a complete Ga surface
layer.

We first calculated the potential energy surface (PES) for Ga and N
adatoms on the (0001) surface, giving immediate insight into
stable sites, migration paths, and diffusion barriers. The potential
energy surface ${E}_{\rm tot}({\bf R}^{\rm ad}_\parallel)$ is calculated
by fixing the adatom laterally at different positions ${\bf R}^{\rm
ad}_{\parallel}$ and allowing all other atoms and the adsorbate height
to relax. A two-dimensional (2D) cut through the resulting potential energy surfaces for
Ga and N adatoms along the energetically lowest path is shown in
Fig. 2. For Ga adatoms we find two adsorption sites: fcc and hcp 
(see Fig. 1), which are degenerate in energy within the accuracy of our 
method. 
Fig. 2 also reveals that the energetically
lowest transition site [corresponding barrier: (0.4 eV)] is the
bridge position; and the migration
over the on-top position is energetically very unfavorable. Actually,
an analysis of the complete PES showed that the on-top position is the
global maximum. 

For N adatoms a qualitatively very different PES results (Fig. 2,
dashed line): While the fcc position is the energetically lowest
binding site, the  hcp position is significantly higher in energy
(1.4 eV). Therefore, to hop from one  fcc site to the next, the N
adatom diffuses along the  bridge position over the hcp
site which is the transition site leading to a barrier of 1.4\,eV.
The large energy difference between the fcc and hcp
sites can be understood by analyzing the atomic structure of the
surface: At the hcp position, the N adatom experiences
directly underneath a N atom in the second layer. Since GaN is a
partly ionic material this leads to electrostatic
repulsion\cite{bernholc97}.

When analyzing adsorption sites it should be kept in mind that the
on-top position is the {\em epitaxial} site for N adatoms on the
(0001) surface, i.e., for {\em high} coverages this site must be a
global minimum. From Fig. 2 we find this site to be energetically
highly unfavorable for an isolated N adatom (i.e. for low
coverages). Therefore, when going from low to high adatom coverages
the N adatoms have to shift from the fcc site (stable at low coverages)
to the on-top position (stable at high coverages). Since this process is kinetically hindered, the 
initial adsorption at the fcc position might be a potential source for stacking
faults.

We have also studied the diffusion on the ($000\bar1$)
surface, which is characterized by a very different atomic
geometry where each surface atom has three dangling bonds compared to
only one at the ($0001$) surface (see Fig. 1).  For Ga adatoms
the PES (see Fig. 3) shows very similar features as found on the
($0001$) surface: The  fcc and hcp sites are
energetically degenerate, the diffusion transition state is at the bridge site
with an again very low energy barrier (0.2\,eV).

For N adatoms we find a qualitative different behavior compared to N
atoms on the (0001) surface (see Fig. 3): $(i)$ N adatoms exhibit the
{\em same} binding sites than Ga adatoms and $(ii)$ the diffusion
barrier is significantly lower [0.9\,eV compared to 1.4\,eV on
(0001)]. The lower diffusion barrier can be understood in terms of the very
different character of the chemical bonds on the (0001) and (000$\bar
1$) surfaces. On the ($0001$) surface (see Fig. 1) the
localized and strongly directed $sp_z$ orbitals favor the formation of strong
Ga-N bonds giving low coordinated sites, as e.g. the bridge position, a rather
high energy. On the other hand, the ($000{\bar 1}$) surface is
characterized by metallic bonds between the surface atoms: the
adsorbate-substrate interaction is hence significantly weaker
resulting in lower diffusion barriers.

An important consequence of our results is that Ga adatoms have a
significantly lower diffusion barrier than N adatoms for both
orientations: Ga adatoms will be orders of magnitude more mobile than
N adatoms at typical growth temperatures. The low diffusion barrier is
a direct consequence of the fact, that for GaN equilibrium surfaces
are Ga stabilized. Thus, for Ga atoms the adsorbate-substrate
interaction is predominantly realized by delocalized metallic Ga-Ga bonds. Since
Ga-bulk melts already slightly above room temperature ($T_{\rm
melt}=30^{\rm o}C$), the Ga-Ga bonds are weak and the adatoms behave
almost like a liquid film on the surface. A similar effect has not
been reported at ``traditional'' III/V semiconductor surfaces like
e.g. GaAs where the surfaces {\it do not} exhibit a metallic like character. The diffusion barrier on these surfaces is thus mainly
characterized by breaking strong cation-anion bonds.  As a consequence
a significantly higher Ga diffusion barrier is found on these
surfaces, e.g., for the polar GaAs\,(111) surface the barrier is 0.9
eV \cite{kley97}.

The significantly larger diffusion barrier for N adatoms compared to
Ga adatoms on equilibrium surfaces has an important consequence:
Although N adatoms on these surfaces are thermodynamically unstable
against evaporation as N$_{2}$-molecules even under N-rich conditions
\cite{neugebauer98}, they can be {\em kinetically} stabilized at the
surface. In order to evaporate, two N atoms have to form a N$_{2}$
molecule. Since migration of N adatoms is a highly activated process
(but necessary to form molecules), the desorption rate may become
smaller than the adsorption rate. Consequently, extended regions in
which the surface is primarily covered by N atoms may be formed. These
{\em N adatoms} are likely to influence the migration path and the
diffusion barrier of Ga adatoms. We have therefore studied the
diffusion of Ga adatoms on the N-terminated ($0001$) and
($000\bar1$) surfaces which can be considered as ``extreme'' cases of
N coverage. The results are shown in Fig 4. The energetically favored
binding sites are located at the fcc and hcp positions and the
transition site is the bridge position. For both surface orientations
the diffusion barrier of Ga adatoms is strongly affected: At ($0001$)
the migration barrier increases from 0.4 to 1.8 eV while at
($000\bar1$) it increases from 0.2 to 1.0 eV. We therefore conclude
that excess N at the surface, which can be formed under N-rich
conditions, significantly {\em reduces} the mobility of Ga
adatoms. The reason is the formation of strong Ga-N bonds, which have
to be broken during the migration of the adatoms.

The reduction in surface diffusivity when going towards more N-rich
conditions is consistent with a recent very detailed MBE growth study
by Tarsa {\it et al.} \cite{Speck97}. We note however, that a direct
comparison between our calculated diffusion barriers and
experimentally derived barriers is by no means straightforward. Experimentally, 
{\em effective} diffusion parameters are commonly derived which average over a 
large area including not only
the clean surface but also steps, impurities, dislocation etc. Since
the effective diffusion barrier is usually dominated by the process
with the {\em highest} barrier, our calculated value (which is for the
clean surface) is a {\em lower} limit of the effective diffusion
barrier. 

The very different mobilities of Ga and N adatoms and the formation of
excess N have important consequences for the growth of GaN. In the
Ga-rich regime where the amount of excess N on the surface is small,
the Ga adatoms are highly mobile and a step-flow mode resulting in 2D
growth is expected. As a consequence, the surface morphology should be
improved and a low density of stacking faults is expected. Further, if 
excess Ga adatoms are present on the surface (as expected for Ga-rich
growth conditions), N adatoms can be efficiently incorporated: The 
probability that fast moving Ga adatoms capture N atoms is much higher than
the other process where N atoms form molecules and desorb from the surface. However,
at the N-terminated surfaces our results show a roughly five times
higher diffusion barrier, indicating that the Ga diffusion length is
significantly shorter under N-rich conditions. Once the diffusion
length is shorter than the mean distance between the binding sites a
statistical roughening of the surface can be expected. Furthermore,
the adatoms might be ``trapped'' at sites not corresponding to the
ideal bulk positions. For example, if a Ga atom is "trapped" at the
fcc position and does not hop to the wurtzite hcp site, a fcc
nucleation center is formed. Thus, the higher adatom mobility under
more Ga-rich conditions will also significantly reduce the density of
stacking faults. Based on a detailed analysis of the adatom kinetics
we expect therefore slightly Ga-rich conditions to be optimal for the
growth of GaN. We note however, that while going towards extreme
Ga-rich conditions might further improve the growth morphology, it
will also enhance the impurity incorporation \cite{tbpublished} and
result in high background carrier concentrations.

We gratefully acknowledge financial support from the BMBF, the Fond
der Chemischen Industrie (T.Z.), and the Deutsche Forschungsgemeinschaft
(J.N.).

\setlength{\unitlength}{1mm}
\begin{figure}[tb]\centering
  \begin{picture}(90,76)(0,-3)
  \psfig{file=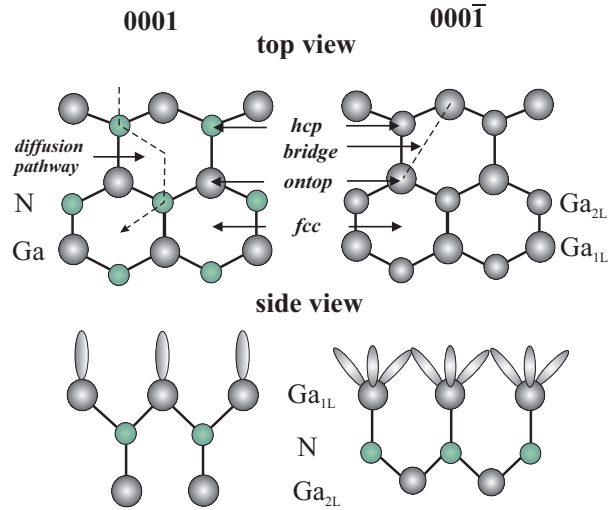,width=8cm,clip=f}
  \end{picture}
\caption{
  Atomic structures (top and side view) for the ($0001$) and ($000\bar1$) Ga
 terminated surfaces. The dashed line in the top left marks the diffusion pathway corresponding to Fig. 2 and Fig. 3. $Ga_{1L}$ marks a Ga-%%@
atom in the first layer, $Ga_{2L}$ in the second.}
\end{figure}

\setlength{\unitlength}{1mm}
\begin{figure}[tb]
 \begin{picture}(70,60)(0,-3)
  \psfig{file=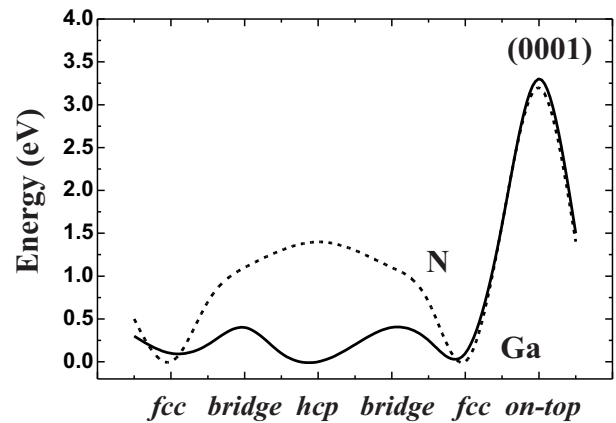,width=8cm,clip=f}
 \end{picture}
\caption{ Total energy (in eV) for Ga-adatom (solid line) and a N-adatom (dashed
  line) at the Ga-terminated ($0001$) surface. The
  energy zero corresponds to the energetically lowest adsorption
  site.}
\end{figure}

\setlength{\unitlength}{1mm}
\begin{figure}[tb]
 \begin{picture}(70,65)(0,-3)
  \psfig{file=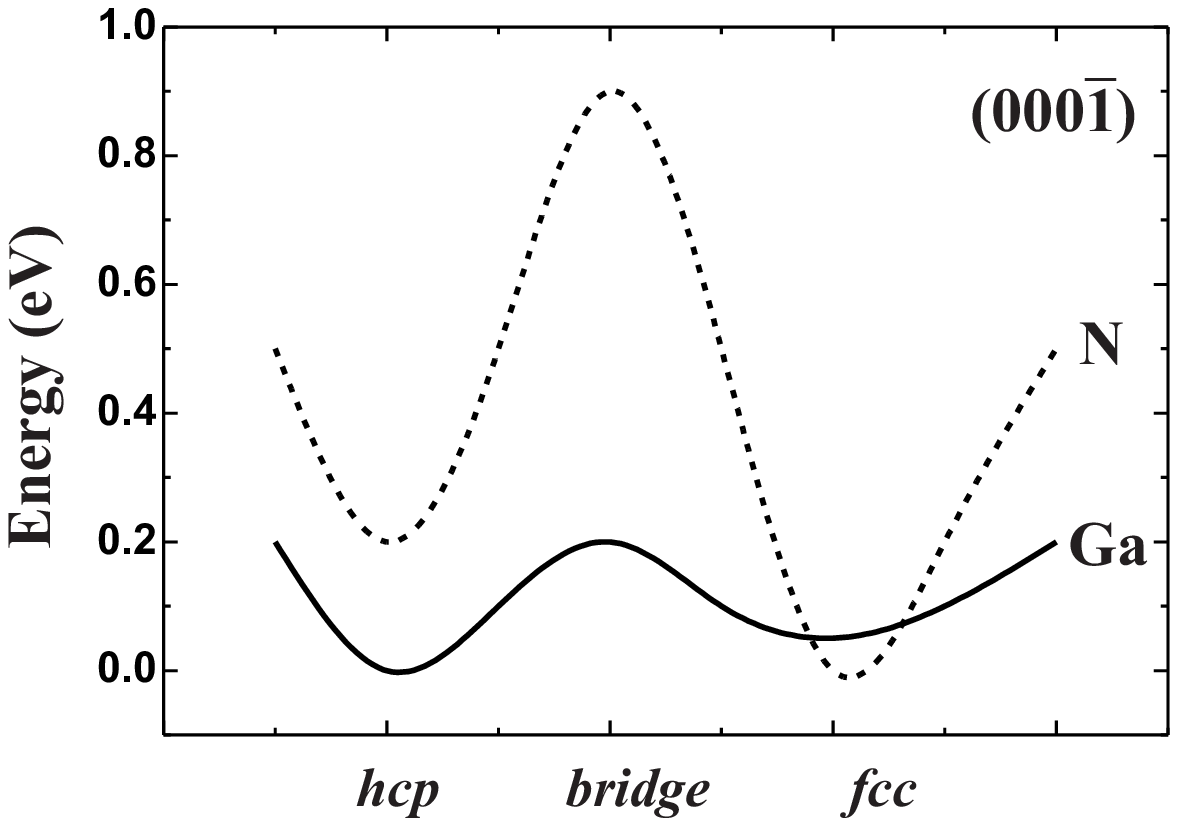,width=8cm,clip=f}
 \end{picture}
\caption{ Total energy (in eV) for a Ga-adatom and a N-adatom at
  the Ga-terminated ($000\bar1$) surface. The energy zero corresponds
  to the energetically lowest adsorption site.}
\end{figure}

\setlength{\unitlength}{1mm}
\begin{figure}[tb]
 \begin{picture}(70,65)(0,-3)
  \psfig{file=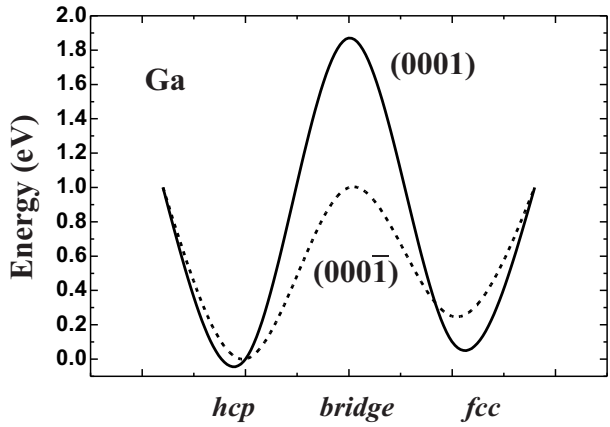,width=8cm,clip=f}
 \end{picture}
\caption{ Total energy (in eV) for Ga-adatoms at the N-terminated ($0001$, solid line) and the N-terminated ($000\bar1$,
  dashed line) surface. The energy zero corresponds to the
  energetically lowest adsorption site.}
\end{figure}

\end{document}